\documentclass[review]{elsarticle}

\usepackage{hyperref}
\usepackage{mathtools, amsfonts,mathrsfs}
\usepackage{subcaption}
\usepackage{graphicx}
\usepackage{booktabs}
\usepackage{threeparttable}
\usepackage[toc]{appendix}
\usepackage{amssymb}
\usepackage{color}
\usepackage{booktabs}
\usepackage{tabularx}
\usepackage{hypernat}
\usepackage{natbib}
\usepackage{tikz}
\usepackage{adjustbox}
\usepackage{multirow}
\usepackage{amsmath}
\usepackage{calc}
\usepackage{accents}

\journal{VSI: Sensitivity analysis 2019}
\newcommand{\Var}{\mathrm{Var}}
\newcommand{\bigcell}[2]{\begin{tabular}{@{}#1@{}}#2\end{tabular}}

\newcommand{\vardbtilde}[1]{\tilde{\raisebox{0pt}[0.85\height]{$\tilde{#1}$}}}
\graphicspath{{Figures/}}

\begin{document}
\numberwithin{equation}{section}
\begin{frontmatter}

\title{Non-intrusive and semi-intrusive uncertainty quantification of a multiscale in-stent restenosis model}

\author[uvaaddress]{Dongwei Ye\corref{mycorrespondingauthor}}
\cortext[mycorrespondingauthor]{Corresponding authors}
\ead{d.ye@uva.nl}

\author[uvaaddress]{Anna Nikishova\corref{mycorrespondingauthor}}
\ead{A.Nikishova@uva.nl}

\author[nlescaddress]{Lourens Veen}
\ead{L.Veen@esciencecenter.nl}

\author[uvaaddress,itmoaddress,emcaddress]{Pavel Zun}
\ead{Pavel.Zun@gmail.com}

\author[uvaaddress]{Alfons~G.~Hoekstra}
\ead{A.G.Hoekstra@uva.nl}

\address[uvaaddress]{Computational Science Lab, Informatics Institute, Faculty of Science, University of Amsterdam, The Netherlands}
\address[nlescaddress]{Netherlands eScience Center, Amsterdam, The Netherlands}
\address[itmoaddress]{ITMO University, Saint Petersburg, Russia}
\address[emcaddress]{Erasmus University Medical Center, Rotterdam, The Netherlands}

\begin{abstract}
The In-Stent Restenosis 2D model is a fully coupled multiscale simulation of post-stenting tissue growth, in which the most costly submodel is the blood flow simulation. This paper presents uncertainty estimations of the response of this model, as obtained by both non-intrusive and semi-intrusive uncertainty quantification. A surrogate model based on Gaussian process regression for non-intrusive uncertainty quantification takes the whole model as a black-box and maps directly the three uncertain inputs to the quantity of interest, the neointimal area. The corresponding uncertain estimates matched the results from quasi-Monte Carlo simulations well. In the semi-intrusive uncertainty quantification, the most expensive submodel is replaced with a surrogate model. We developed a surrogate model for the blood flow simulation by using a convolutional neural network. The semi-intrusive method with the new surrogate model offered efficient estimates of uncertainty and sensitivity while keeping a relatively high accuracy. It outperformed the results obtained with earlier surrogate models. It also achieved the estimates comparable to the non-intrusive method with a similar efficiency. Presented results on uncertainty propagation with non-intrusive and semi-intrusive metamodelling methods allow us to draw some conclusions on the advantages and limitations of these methods.

\end{abstract}


\end{frontmatter}


\section{Introduction}
Numerical simulations of real-world phenomena contribute to a better understanding of these phenomena and to predicting the dynamics of the underlying systems. Many natural phenomena occur across scales in space and time \cite{hoekstra2014multiscale,groen2014survey,Mizeranschi2016, Chopard2018,alfons2019}. As a result, multiscale models and simulations are widely used \cite{Praprotnik_2008, sloot2009multi,hoekstra2014multiscale,karabasov2014multiscale,alowayyed2016multiscale}. These multiscale models couple mathematical models of relevant processes on different spatial or temporal scales together using suitable scale bridging methods \cite{chopard2014framework}. 
However, multiscale simulations can suffer from substantial computational cost because of the high computational demands of, usually, the microscale simulations \cite{alowayyed2016multiscale}. Uncertainty quantification (UQ) analysis \cite{smith2013uncertainty} applied to multiscale simulations adds additional substantial computational burden since thousands of runs are required for good estimates of the uncertainties. Therefore simulations can become extremely time-consuming or impractical, even on current state-of-the-art supercomputing infrastructure.

Surrogate modelling is a common non-intrusive way to resolve the computational intensity problem under the UQ scenario. The motivation for this technique stems from the large number of samples needed in the UQ estimates or from frequently required numerical integration in global sensitivity analysis with the Monte Carlo method \cite{fang2005design}. A non-intrusive surrogate model takes the complete model as a black box, mimics the behaviour of the original computational model from a limited amount of existing data and evaluates the corresponding responses of the model based on specific inputs \cite{forrester2008engineering}. By replacing the original model with a surrogate model, this allows conducting UQ or sensitivity analysis at an acceptable computational cost. Gaussian process regression \cite{Rasmussen2005,STAUM2009} is one of the state of art methods for surrogate modelling. It is a non-parametric Bayesian regression method and its predictive distribution can be efficiently used in UQ and sensitivity analysis \cite{OAKLEY2004,MARREL20084731,CONTI2010640,GRATIET2014,BILIONIS20125718}. There are several other popular methods for surrogate modelling, including polynomial chaos expansion  \cite{SFEM2006,SUDRET2008964} and neural networks \cite{TRIPATHY2018565,EASON2014220}.

As opposed to the traditional non-intrusive UQ method which takes the whole simulation as a black box, we have recently proposed a set of semi-intrusive algorithms for multiscale UQ \cite{nikishova2019semi} and demonstrated their effectiveness for several multiscale UQ scenarios \cite{nikishova2019semi,Nikishova_2018_Semi-intrusive}. The term "semi-intrusive" refers to additional interventions into the code of the model compared to non-intrusive approaches: one "opens up" the black box and considers the coupled structure of the multiscale model while the single scale models are still viewed as black boxes. Usually the output of a multiscale model is derived from a macroscale submodel, which in turn is implicitly determined by microscale dynamics to which it is coupled. One approach 
from \cite{nikishova2019semi} relies on performing a Monte Carlo UQ on the macroscale submodel while replacing the most costly microscale submodel by a surrogate model. Replacing the expensive part of a model with a relatively cheap surrogate can often significantly improve computational efficiency, but comes at the cost of reduced accuracy. 

In \cite{Nikishova_2018_Semi-intrusive}, a physics-based and an interpolation-based surrogate model were constructed to implement the semi-intrusive UQ for the in-stent restenosis multiscale model \cite{evans2008application,Tahir2011,tahir2013modelling}. In that research, the flow solver submodel was replaced with these surrogates. The flow solver takes blood vessel geometry and blood flow velocity as the inputs and outputs the corresponding wall shear stress. The physics-based surrogate model simplified the fluid simulation to an ideal Poiseuille flow, hence the shear stress could be computed analytically. The interpolation-based surrogate model applied the nearest neighbour method and approximated the shear stress based on a training dataset. We estimated how the three uncertain inputs: blood flow velocity, endothelium regeneration time and maximum deployment depth of the stent, contributed to the quantity of interest (QoI), the neointimal area. All the semi-intrusive UQ were implemented with a quasi Monte Carlo method based on a Sobol sequence \cite{SOBOL1976236,KUCHERENKO2009}. The results were compared to black-box Monte Carlo results to demonstrate the efficiency improvement. However, a comparison between the semi-intrusive algorithm and non-intrusive UQ with a surrogate model replacing the complete multiscale model were not explored in that study. In this work, we first improve the surrogate model of micro-scale model (the blood flow simulation) from \cite{Nikishova_2018_Semi-intrusive} by using a convolutional neural network (CNN), because of its capability of pattern recognition and feature extraction \cite{Guo:2016:CNN:2939672.2939738,CNNonFEA}. Additionally, a surrogate model for non-intrusive UQ is designed to directly map the input parameters to the quantity of interest. The uncertainty estimations with both these methods are then carried out and compared in terms of the estimation accuracy and computational efficiency. 

The paper is arranged as follows. The two-dimensional multiscale model of in-stent restenosis is shortly introduced in Section~\ref{sec:model}. The new surrogate model for the blood flow simulation and a surrogate model for the whole in-stent restenosis model are described in Section~\ref{sec:metamodel} and \ref{sec:GRPsurrogate}. The approach to estimating and analyzing the uncertainty of the response of the in-stent restenosis model is explained in Section~\ref{sec:uq_sa}. The results of surrogate modelling, uncertainty quantification and sensitivity analysis are presented in Section~\ref{sec:results}. Sections~\ref{sec:Discussion} and~\ref{sec:Conclusion} compare and discuss the UQ performance and summarise the obtained results.



\section{Model}\label{sec:model}
\subsection{In-stent Restenosis}

An arterial stenosis is the abnormal narrowing of an artery, usually due to accumulation of fatty material in the walls and intimal thickening (atherosclerosis). In ischemic heart disease, a stenosis in a coronary artery limits blood flow to the heart muscle, which can result in reduced heart function, shortness of breath, chest pains or a heart attack. Coronary stenosis can be treated using balloon angioplasty, in which a balloon is inserted into the artery via a catheter and inflated, which compresses the fatty plaque against the arterial wall. During this procedure, a wire-mesh stent is deployed to keep the artery from recoiling back to the narrowed state. This procedure damages the vessel wall, and in particular the endothelium, the innermost lining of the artery. This triggers a healing response involving (amongst other processes) growth and proliferation of smooth muscle cells (SMCs) on the inside of the artery. In some cases, an excessive growth response occurs, leading to a significant renewed narrowing of the artery inside of the stent. This is known as an in-stent restenosis (ISR), and is considered an adverse treatment outcome \cite{Jukema2012, Jukema2012a, Iqbal2013a}.

\begin{figure}[tb]
\centering
\includegraphics[width=0.9\textwidth]{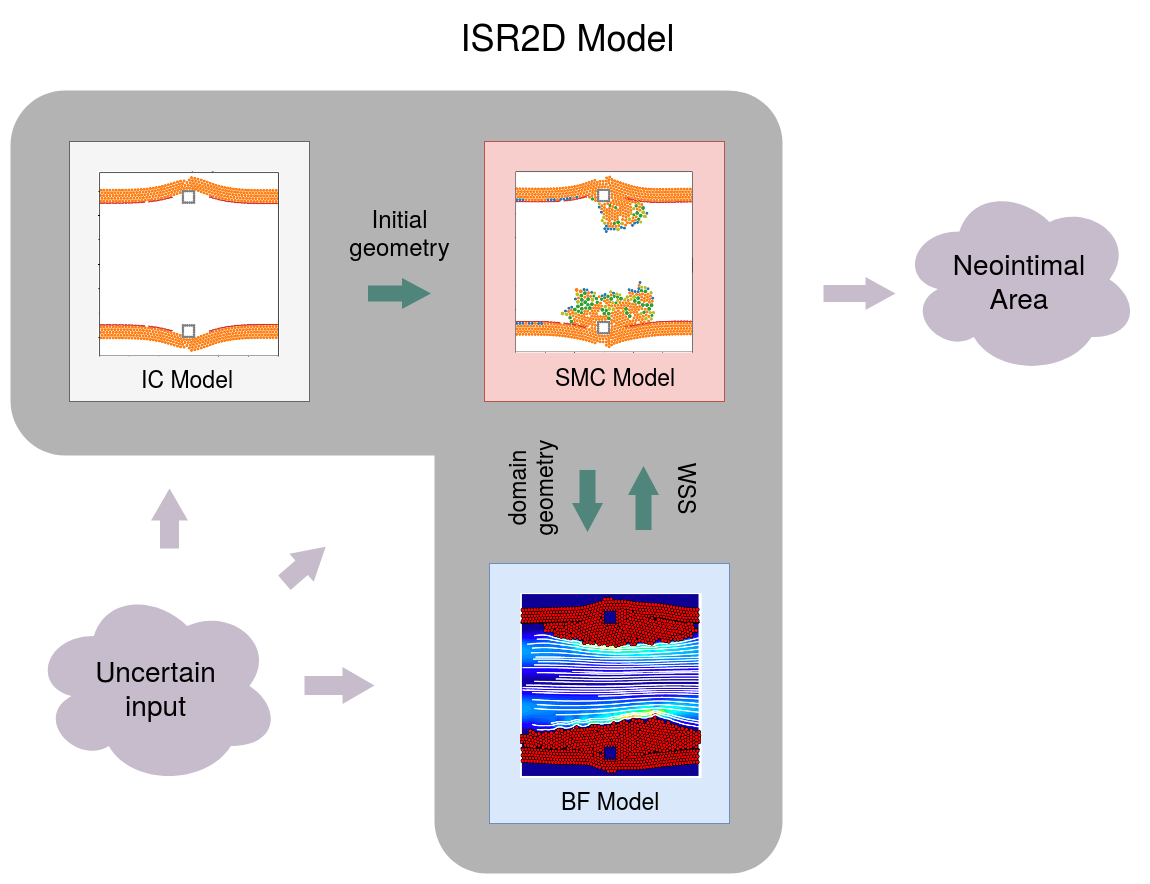}
\caption{Diagram of the ISR2D model: the initial conditions (IC) submodel provides the initial geometry to the smooth muscle cells (SMC) single scale model where the geometry is further updated. SMC model calls the blood flow (BF) simulation, which provides the wall shear stress (WSS) for the updated geometry. The cycle continues until the final time step is reached, when the SMC model yields the final output of the cross-sectional area.
}
\label{fig:isr2d_struc}
\end{figure}

The ISR2D model is a two-dimensional simulation of the post-stenting healing response of an artery \cite{Tahir2011,tahir2013modelling}, which is used here to test the proposed semi-intrusive multiscale UQ algorithm. Note that a more realistic, but also computationally much more expensive three dimensional version of the model is available \cite{Zun2017,Zun2019}. The ISR2D model used in this paper consists of three submodels: the IC submodel, which simulates initial conditions in the form of the state of the artery immediately after stenting, the SMC submodel, which is an agent-based simulation of smooth muscle cell growth and endothelium recovery, and the blood flow submodel, which uses the Lattice Boltzmann method (LBM) to simulate blood flow through the artery. The structure of ISR2D is shown in Figure \ref{fig:isr2d_struc}. 

Sufficiently high wall shear stress (WSS) at the arterial wall triggers any present endothelium to produce nitric oxide, which in turn inhibits the growth of the SMCs if it crosses a threshold value. Blood flow thus affects SMC growth, but in turn is also affected by it, as the proliferating SMCs change the geometry of the artery. Note that due to random placement of daughter cells when the growing SMCs divide, variability in the length of the cell cycle, and a random spatial pattern of endothelium recovery, the SMC model is stochastic. The main output of the model is the cross-sectional area of the neointima (the new tissue formed due to SMC proliferation) as a function of time after stenting. A clinically recognised in-stent restenosis occurs if more than 50\% of the original cross-sectional area of the artery is covered by the neointima \cite{Serruys2009}.

The SMC growth occurs over a period of weeks, while blood flow adapts much more rapidly to the changing geometry. ISR2D is therefore a multiscale model exhibiting time-scale separation. For every time step of the SMC model (one time step simulates one hour resulting in 1440 steps per 60 days of the total simulation time), the BF simulation is run to convergence for the current geometry, and the resulting WSS values are sent back to the SMC model, which uses those in the model of nitric oxide production by endothelial cells.

\begin{figure}[tb]
\centering
 \includegraphics[width=0.8\textwidth]{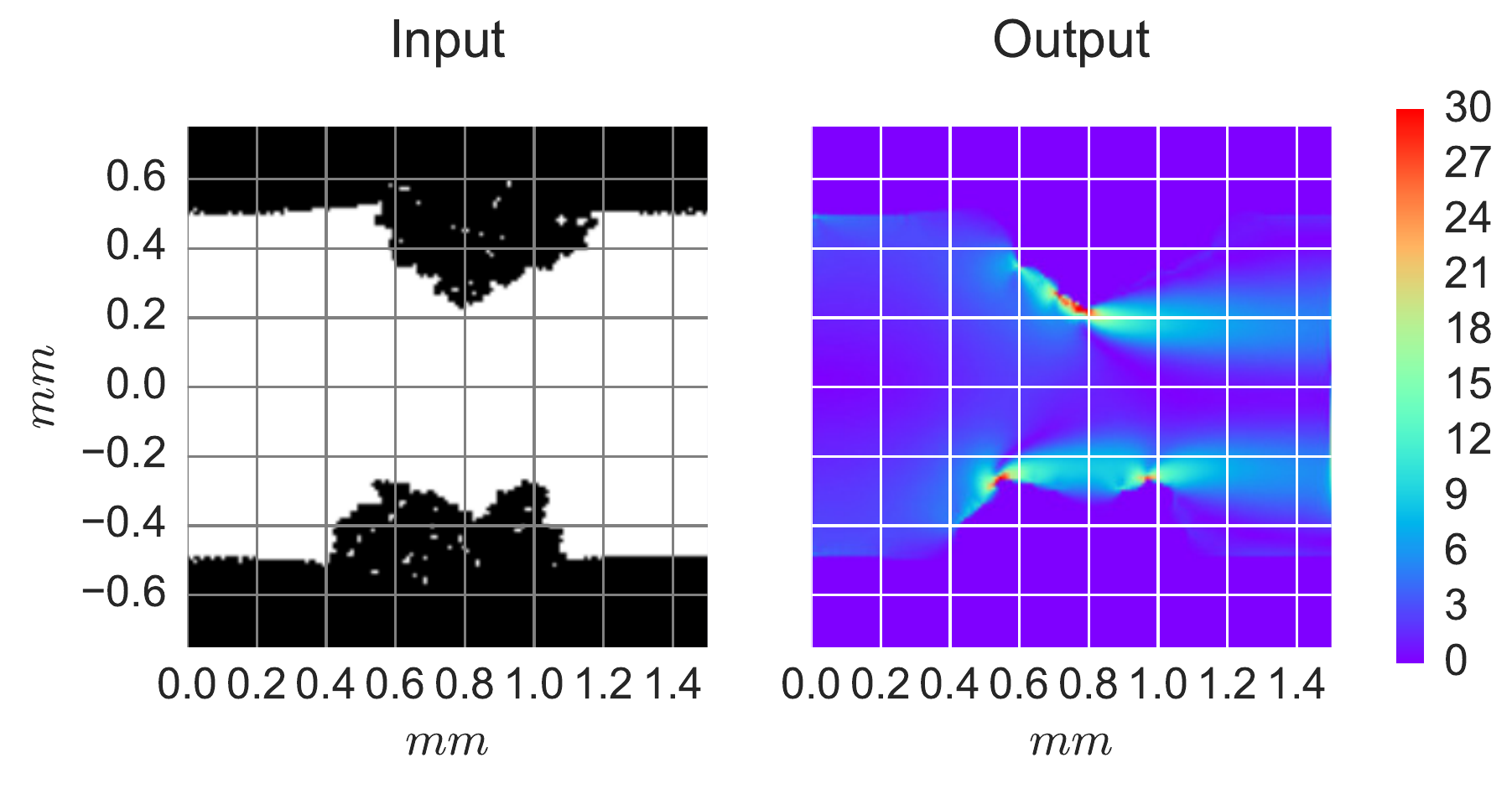}
 \caption{Input and output of blood flow simulation. Left plot: $150 \times 150$ binary geometry map as the input for blood flow simulation. The black part is the vessel wall and the white part is the lumen (the fluid domain). Right plot: simulated shear stress in the domain (in $Pa$).}
\label{fig:bf_input_output}
\end{figure}

Figure \ref{fig:bf_input_output} shows an input (domain map) and corresponding output (shear stress field) of the BF submodel. The domain map represents the geometry of the artery in the form of a binary grid, with 1 (shown in black) representing a solid grid cell, and 0 (shown in white) representing a fluid grid cell. The output of the BF simulation is a corresponding grid of shear stress values, which are set to 0 for solid domain cells, and set to the shear stress values computed by LBM for fluid domain cells. The shear stresses at the fluid-solid boundary layer are taken as the wall shear stresses and passed to the SMC model.

The blood flow simulation is the computationally most expensive component of the ISR2D model. It takes around 80\% of the computational time, and the potential gain in performance obtained by replacing it with a surrogate model in the semi-intrusive UQ scenario is therefore highest. Nikishova et al. \cite{Nikishova_2018_Semi-intrusive} performed a semi-intrusive UQ analysis for the same model, comparing surrogates based on nearest-neighbour interpolation and on simplified physics to a non-intrusive black box quasi Monte Carlo approach. The UQ estimate with physics surrogate has improved computational efficiency but the means of the cross-sectional area of the neointima resulting from the surrogate models were substantially lower than the black-box Monte Carlo results. On the other hand, the uncertainty estimates with nearest-neighbour interpolation were better but the corresponding speedup was relatively poor.
In this paper, an accurate surrogate model using a convolutional neural network is proposed and demonstrated to be a good representation of the blood flow model, while at the same time leading to the desired reduction in computational cost for the multiscale UQ.

\subsection{Surrogate Model for Blood Flow Simulation}\label{sec:metamodel}
As mentioned in the previous section, the blood flow simulation in the ISR2D model is computationally expensive. To reduce the computational cost of the model, a surrogate model to compute the required wall shear stresses is used to replace the original blood flow model. 
Convolutional neural networks have been applied to fetch the features from irregular geometries in fluid dynamics prediction \cite{Guo:2016:CNN:2939672.2939738,Tompson:2017:AEF:3305890.3306035,CNNonFEA}. In the ISR2D application, the aim is to learn the wall shear stress as a nonlinear function of the vessel wall geometry and the blood flow velocity. 

The mapping between input and output of the BF simulation can be considered as a function $f$, which takes the geometry matrix $\boldsymbol{\zeta}$ and the inlet blood velocity $v_{\text{in}}$ as input and produces a $2 \times k$-dimensional vector of WSS magnitudes, $\tau_{\text{wss}}$ as the output:

\begin{equation}
    \tau_{wss} = f(\boldsymbol{\zeta},v_{\text{in}}),
\end{equation}
where $k = 150$ is the grid size along x axis, $\boldsymbol{\zeta} = (\zeta_{ij}) \in \mathbb{R}^{k \times k}$. The geometry matrix $\boldsymbol{\zeta}$ was used for CFD simulation. The surrogate model $\hat{f}$ replaces the original blood flow model $f(\boldsymbol{\zeta},v_{\text{in}})$ and offers an approximate prediction of wall shear stress in a reduced amount of time. 

\begin{figure}[tb!]
\centering
\includegraphics[width=\textwidth]{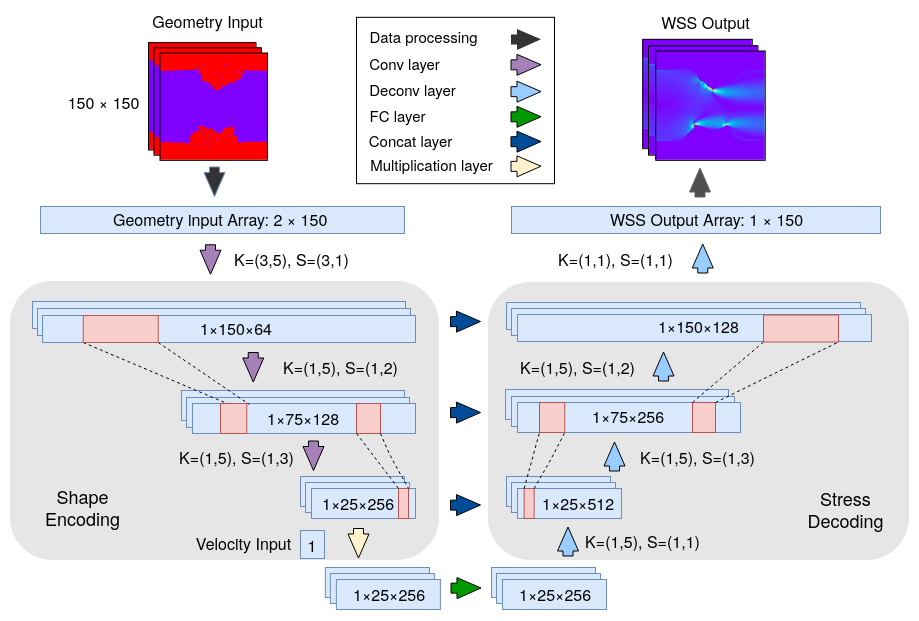}
\caption{Diagram of the CNN model. Arrows with different colors denotes different kind of operation layers. The numbers in each light blue box denote the dimensions of the output after each operation layer (Height $\times$ Width $\times$ Feature maps). K and S denote the size of kernel and the stride in each convolution and deconvolution layer. Padding setting 'same' is applied to remove the influence of kernel size on the dimension of output in each convolution and deconvolution layer.}
\label{fig:CNN_struc}
\end{figure}

The CNN model follows the network structure proposed by \cite{Guo:2016:CNN:2939672.2939738} and was optimized to fit our application. The model consists of three parts: shape encoding, nonlinear mapping and stress decoding, as shown in Figure~\ref{fig:CNN_struc}. The shape encoding layers extract the features of the geometry to the shape code. A fully connected (FC) layer then maps the shape code together with the blood flow velocity to the stress code. The stress decoding part is responsible for a mapping from the stress code to wall shear stress. In this surrogate model, the geometry input was transformed from a binary map to a $2 \times k$ array which indicates the locations of upper and lower fluid-solid boundaries. The convolution layers then take the information from both boundaries into account and predict the shear stress on these boundaries. There are three convolution layers, a fully connected layer and four deconvolution layers deployed between the input layer and output layer. Each of them is followed by a rectifier linear unit (ReLU) as the activation function. Additionally, the output of each convolution layer is concatenated to the corresponding deconvolution layer to help with the decoding process. The output of the surrogate model represents the shear stresses on the wall surface on one side of the channel. The sequence of the two rows in the input decides which side is predicted. Therefore, a prediction of the WSS on both boundaries requires the surrogate model to be run twice.

The training data for the surrogate model comes from the runs of the ISR2D model \cite{Nikishova2018}. One run of ISR2D calls the LBM solver 1440 times (once per hour of simulated time). This means that with only a few runs of the simulation, a considerable amount of flow data for training is already available. We trained the surrogate model with the data from four runs of the ISR2D simulation, hence 5760 blood vessel geometries and wall shear stress distributions were used for training. A validation dataset was generated from an additional run of the ISR2D simulation. These five runs of simulation for training and validation are the results from previous experiments using the quasi Monte Carlo method \cite{Nikishova2018}. The training optimization was based on the mean squared error loss function: 


\begin{equation}
    \mathcal{L}(\tau_{wss}) = \frac{1}{n} \sum_{i = 1}^{n} \|\tau_{wss}^{(i)}- \hat{\tau}_{wss}^{(i)} \|^2,
\end{equation}
where $n$ denotes the number of samples of the training set. The \textit{Adam} optimizer \cite{kingma2017adam} was used to optimize the model parameters. A test against the validation dataset was carried out during the training process to prevent the model from overfitting. The epoch was set to 80 as the loss does not decrease significantly after that. The surrogate model was implemented in Keras \cite{chollet2015keras}.

\subsection{Surrogate Model for ISR2D}\label{sec:GRPsurrogate}
This surrogate model is constructed to replace the whole ISR2D multiscale model for uncertainty quantification and sensitivity analysis. Let the multiscale model function be defined by $g(\xi)$ with $\xi$ denoting an $n$-dimensional vector consisting of the stochastic variables and uncertain inputs of the model. The response of the model is:

\begin{equation}
    y = g(\xi).
\end{equation}
As mentioned before, the ISR2D model is a stochastic model and thus includes both aleatory uncertainty and epistemic uncertainty. In the surrogate model for non-intrusive UQ, we assume that the aleatory uncertainty can be separated from the function $g$, hence the expression for the surrogate model can be written as:

\begin{equation}
    \hat{y} = h(\xi_{\sim \xi^*}) + \epsilon(\xi^{*}), 
\end{equation}
where $\hat{y}$ denotes the response of the surrogate model, $\xi^*$ is the random variable containing aleatory uncertainty and $\xi_{\sim \xi^*}$ is the parameter vector without stochastic variable. Assuming that the stochasticity $\epsilon$ follows a normal distribution $N(0,{\sigma^{*}}^2)$, such a stochastic model can be quantified by Gaussian process regression~(GPR). Note that ISR2D is a time-evolving model, and that we are not only interested in the response at the final timestep, but also in the dynamics of the process. To avoid an extra dimension of input, more precisely, a time $t$ that will significantly increase the computational cost of training and prediction \cite{Rasmussen2005}, a local surrogate model for each time step is constructed. An alternative choice could be to apply sparse heteroscedastic GP \cite{SparseGP2006,HGP2011} and to adopt the time as an extra dimension of input. However, this would require additional estimations for the variances of local input noises and pseudo-inputs for the sparse Gaussian process in the marginal likelihood optimisation. Such high dimensional optimisation may result in a relatively poor inference of the hyperparameters. Therefore we adopt the local surrogate models, and the expression can be rewritten as:

\begin{equation}
    \hat{y}_t = h_t(\xi_{\sim \xi^*}) + \epsilon_t(\xi^{*}), \; t=1,2,...,T.
\end{equation}
GPR is based on the assumption of the joint Gaussian distribution between training data and prediction mean:

\begin{equation}
    \begin{pmatrix} y_t^{\text{train}} \\ h_t^{\text{pred}} \end{pmatrix}  \sim N\begin{pmatrix} 0, \begin{bmatrix} K_t + {\sigma_t^{*}}^2 I & \tilde{K_{t}} \\ (\tilde{K_{t}})^T & \vardbtilde{K_t} \end{bmatrix}    \end{pmatrix}, \; t = 1,2,...,T, 
\end{equation}
where covariance matrices $K_t$, $\vardbtilde{K_t}$ and $\tilde{K_t}$ denote the correlation within training data, within new data, and between these two respectively at each time step. Applying Bayesian inference, the posterior probability  distribution of $h_t^{\text{pred}}$ given training set$(\xi_{\sim\xi^*}^{\text{train}},y_t^{\text{train}})$ follows a Gaussian process:

\begin{equation}
    \label{eq:Bayesian_inference}
    P(h_t^{\text{pred}} |  (\xi_{\sim \xi^*}^{\text{train}},y_t^{\text{train}}),\xi_{\sim \xi^*}^{\text{pred}}) =  \mathcal{GP}( \bar{h}_t^{\text{pred}},\text{Var}({h_t^{\text{pred}}})),
\end{equation}
\noindent where 
\begin{equation*}
    \bar{h}_t^{\text{pred}} = \tilde{K_t} (K_t+{\sigma_t^{*}}^2 I)^{-1} y_t^{\text{train}},
\end{equation*}
\noindent 
\begin{equation*}
    \text{Var}({h_t^{\text{pred}}}) = \vardbtilde{K_t} - \tilde{K_t} (K_t+{\sigma_t^{*}}^2 I)^{-1} (\tilde{K_t})^T.
\end{equation*}

\noindent The mean value offers the prediction and the variance represents the uncertainty of this prediction. The radial basis function kernel was chosen for the computation of covariance matrices:

\begin{equation}
    k_t\left(x_{i}, x_{j}\right)= \sigma_t^f  \exp \left(-\frac{ \|x_{i}-x_{j}\|^2}{2 l_t^{2}}\right),
\end{equation}
where $\|x_i-x_j\|$ denotes the $L^2$ norm between two points in the Gaussian process. In each local GPR model, there are three hyperparameters associated with the kernel: length scale $l_t$, signal variance $\sigma_t^f$ and noise variance (stochasticity) $\sigma_t^{*}$. These hyperparameters are trained by optimizing the log marginal likelihood function:

\begin{equation}
\begin{split}
    \log p(y_t^{\text{train}}|\xi_{\sim \xi^*}) = & -\frac{1}{2} (y_t^{\text{train}})^{\top} (K_t + \sigma_t^{*2} I)^{-1} y_t^{\text{train}} \\  &- \frac{1}{2} \log |K_t + \sigma_t^{*2} I| - \frac{n}{2} \log 2\pi.
\end{split}
\end{equation}
The training data comes from the qMC result from \cite{Nikishova2018} and the size of the training data was chosen to match the speedup of semi-intrusive UQ. The details of speedup calculation are introduced in the Section~\ref{sec:speedup}. As the semi-intrusive method gains a speedup of around 7 times, the speedup of the non-intrusive method is also designed to be around 7 for comparison. Therefore, the results of 150 ISR2D simulations from previous experiments using the quasi Monte Carlo method \cite{Nikishova2018} are used for training. The Gaussian process surrogate model used in non-intrusive UQ was built using GPy \cite{gpy2014}.

\section{Methods}\label{sec:uq_sa}

\subsection{Uncertainty quantification and sensitivity analysis}
Uncertainty Quantification is the analysis of uncertainty of a computational model \cite{OBERKAMPF2002333}, including uncertainty in the model response (\textit{forward problem}), and its input (\textit{inverse problem}). Sensitivity analysis (SA) is an important part of UQ, that recognises the effects of each source of uncertainty on the model response variability. Here, the forward propagation of uncertainty from inputs to the output of the ISR2D model together with SA are studied. Three epistemic uncertain inputs are considered in the estimation: blood flow velocity, endothelium regeneration time and stent deployment depth. The blood flow velocity is an input to the BF simulation, the endothelium regeneration time is an uncertain parameter for the SMC model and the deployment depth affects the computation in the IC model.
The ranges of the uncertain parameters are shown in Table~\ref{tab:uncertain_para} in Section~\ref{sec:results}. 

Since ISR2D is a stochastic model, the estimates of the response variance and partial variances in SA include the aleatory uncertainty as well. In other words, the model stochasticity is treated as another uncertainty source similar to the uncertain inputs. This approach was adopted in our previous work \cite[p. 764]{Nikishova2018}. It is important to note that there exist alternative approaches to deal with stochasticity in uncertainty and sensitivity analysis, for instance~\cite{IOOSS20091194, zhu2020global}.

The semi-intrusive metamodelling method involves replacing a computationally expensive single-scale submodel with a surrogate which produces an approximation to the original single scale model result in a reduced time. The input uncertainty is propagated via the surrogate model in the same way as for the non-intrusive method: an ensemble of model outputs is obtained by running the model with different values of the uncertain parameters sampled according to their distributions. Using the obtained samples of the model response, uncertainty in the model is estimated, analysing the probability density function, mean and variance, as well as estimating the Sobol sensitivity indices \cite{Sob90,KUCHERENKO2009}. The mean and the variance of the model responses at time $t$ are then estimated by:

\begin{align}
\begin{split}
 \label{eq:uq_qmc_macro}
 \mathbb{E} \left(\hat{y}_t(\xi)\right) &\approx \frac{1}{N} \sum^N_{j=1} \hat{y}_t(\xi_j),\\
 \Var \left(\hat{y}_t(\xi)\right) &\approx \frac{1}{N} \sum^N_{j=1} \hat{y}_t(\xi_j)^2 - \left(\frac{1}{N} \sum^N_{j=1} \hat{y}_t(\xi_j) \right)^2, 
\end{split} 
\end{align}
where $\hat{y}_t(\xi_j)$ is the value of the model response obtained with the $j$-th sampled value of the uncertain inputs $\xi_j$ and $N$ is the total number of samples. The total Sobol sensitivity index for the $i$-th parameter together with the stochastic parameter $\xi^*$ is defined by:

\begin{align}
\begin{split}
\label{eq:SI_tot}
S^{total}_{\xi_i, \xi^*} = \frac{\Var^{total}_{\xi_i, \xi^*}}{\Var\left(\hat{y}_t(\xi)\right)}, 
\end{split}
\end{align}
where the partial variance in the numerator is approximated by \cite{HOMMA19961, Sal101}:

\begin{align}
\begin{split}
\label{eq:partial_variance_tot}
\Var^{total}_{\xi_i, \xi^*} &\approx \frac{1}{2N} \sum_{j=1}^N {\left(\hat{y}_t(\mathbf{\xi}_j) - \hat{y}_t\left( (\mathbf{\xi}_{\sim \xi_i, \xi^*})_j, (\xi_i)_{j+N}, \xi^*_{j+N}\right)\right)^2},
\end{split}
\end{align}
where $\mathbf{\xi}_{\sim \xi_i, \xi^*}$ is a vector of all uncertain parameters in $\xi$ except $\xi_i$ and the stochastic parameter $\xi^*$, and $\hat{y}_t\left( (\mathbf{\xi}_{\sim \xi_i, \xi^*})_j, (\xi_i)_{j+N}, \xi^*_{j+N}\right)$ denotes the model response at time $t$ with the same values of all inputs as for $\hat{y}_t(\mathbf{\xi}_j)$ except of the $i$-th input $\xi_i$ and of the stochastic parameter $\xi^*$. Since the interpretation of the total sensitivity indices is the portion of uncertainty that would remain if all other parameters were known precisely \cite{LOPIANO2021107300} and the stochasticity is the irreducible part of uncertainty, the effect of stochastic uncertainty is included in the estimator in \eqref{eq:partial_variance_tot}. 




\subsection{Speedup}\label{sec:speedup}
The main purpose of applying SI and NI methods is to speed up the simulation and reduce the computational cost while getting accurate enough estimates of the uncertainties. The speedup of the UQ analysis of these advanced methods was computed as follows:

\begin{equation}\label{eq:speedup}
    \mathcal{S} = \frac{N  \mathcal{T}_{\text{ISR}}}{N\mathcal{T}_{\text{ISR}^*}+\mathcal{T}_{\text{train}}+\mathcal{T}_{\text{sample}}},
\end{equation}
where $N$ is the number of runs of ISR2D simulation in the UQ analysis. $\mathcal{T}_{\text{ISR}}$ is the execution time of an ISR2D simulation with the LBM solver. $\mathcal{T}_{\text{ISR}^*}$ is the execution time of either an ISR2D simulation with a blood flow surrogate model or the execution time of the surrogate model for the non-intrusive method. $\mathcal{T}_{\text{train}}$ denotes the training time for the surrogate model. $\mathcal{T}_{\text{sample}}$ denotes the time to generate the data necessary for the training process.

\section{Results}\label{sec:results}

First, the quality of the blood flow surrogate model is evaluated. Then, the results of UQ and sensitivity analysis based on non-intrusive and semi-intrusive methods are compared to a previously obtained reference solution reported in \cite{Nikishova2018}. 

\subsection{Blood flow surrogate model}
\begin{figure}[htbp]
\centering
 \includegraphics[width=0.85\textwidth]{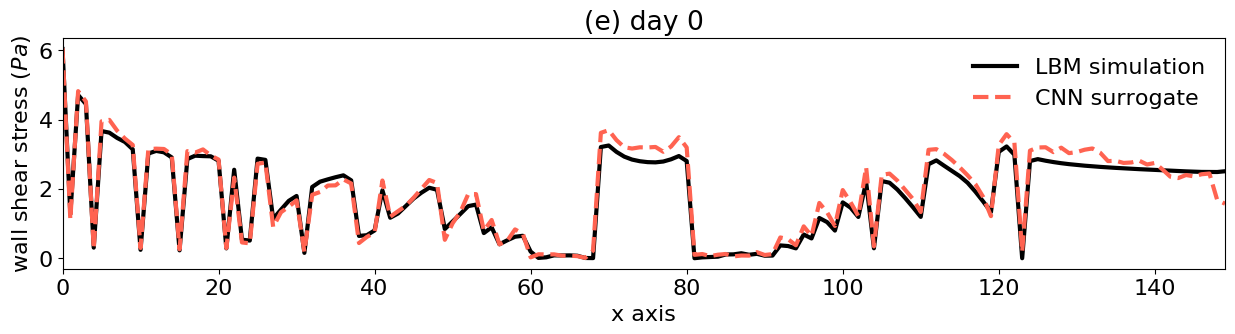}
 \includegraphics[width=0.85\textwidth]{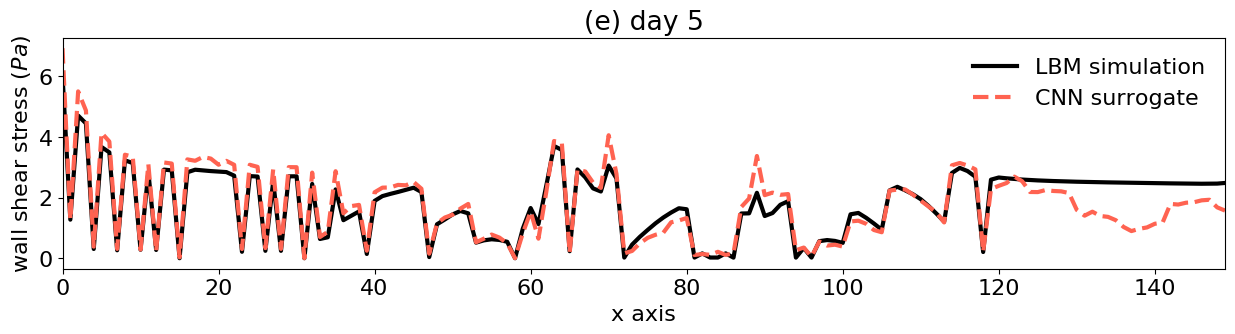}
 \includegraphics[width=0.85\textwidth]{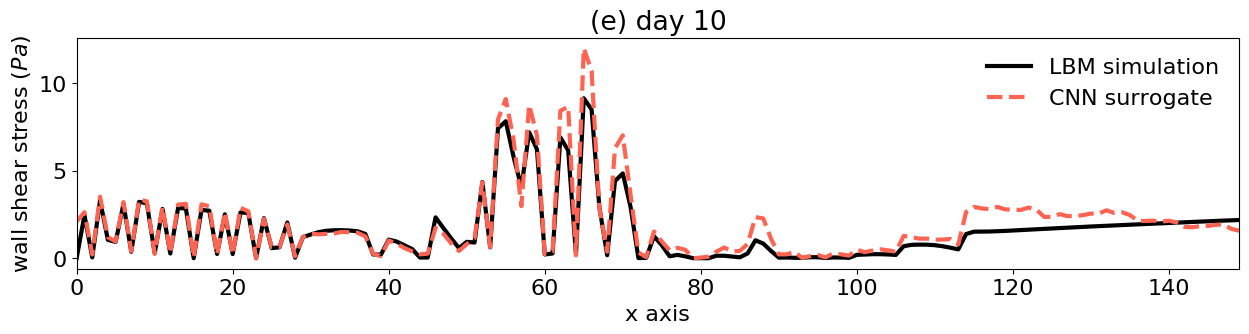}
 \includegraphics[width=0.85\textwidth]{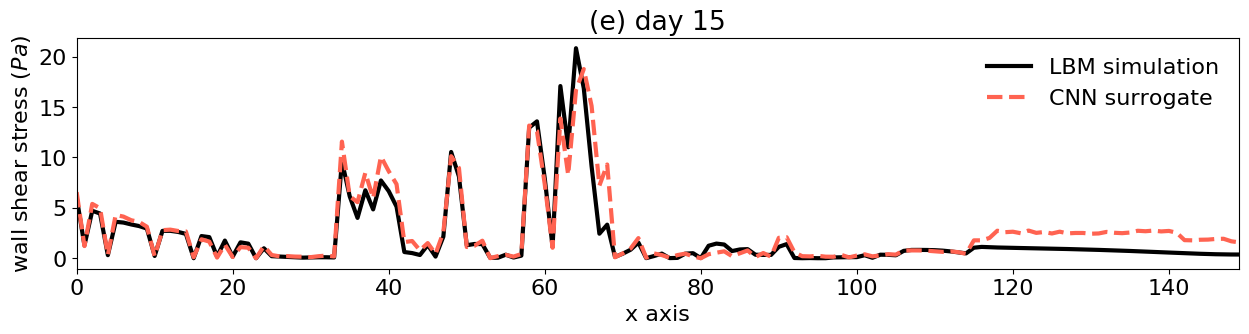}
 \includegraphics[width=0.85\textwidth]{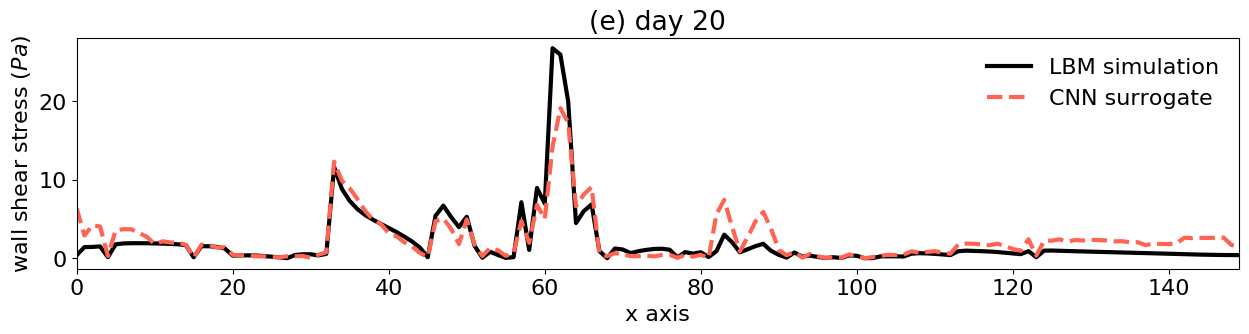}
 \includegraphics[width=0.85\textwidth]{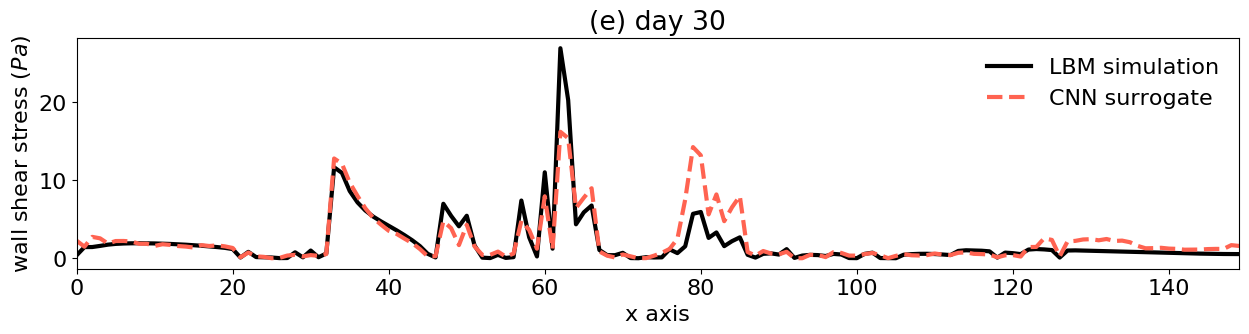}
 \caption{(a) - (f) Predictions of wall shear stress distribution along the upper boundary by LBM and CNN surrogate models at 0, 5, 10, 15, 20, 30 days, respectively. The corresponding NMAEs are 3.16\%, 5.51\%, 4.92\%, 3.92\%, 3.95\% and 3.32\%}
\label{fig:surrogate_compare}
\end{figure}

Before applying the blood flow surrogate model to semi-intrusive UQ analysis, the quality of the surrogate model was evaluated. We used normalized mean absolute error (NMAE) on a test dataset to evaluate the quality of the surrogate model: 

\begin{equation}
    \text{NMAE} = \frac{ \frac{1}{2k}\| \tau_{wss}-\hat{\tau}_{wss}\|}{max\{|\tau_{wss}|\}} \times 100\%,
\end{equation}

\noindent where $\| \cdot \|$ denotes $L_1$ norm and $max\{|\tau_{wss}|\}$ is the peak stress in the LBM. The test dataset was generated from a single additional run of the ISR2D simulation in addition to the training dataset and the validation dataset. Note that the test dataset is not considered in calculating the computational cost of the training process since the test data is not used for training the model. Here, the test dataset is used only to illustrate the performance of the surrogate model. Figure~\ref{fig:surrogate_compare} visualizes the wall shear stress prediction of the CNN surrogate model and the LBM solver on the test dataset. 
The averaged NMAE of the surrogate model on the whole test dataset is around $3.63\%$. The results show that the CNN surrogate model approximates the wall shear stress well in most cases. The prediction gets slightly worse close to the outlet of blood flow. The relatively poor prediction may be caused by extrapolation, since the growth is stochastic, and the lumen geometry may end up with a previously unseen irregular shape which is not covered by the training dataset. 

\subsection{Uncertainty quantification and sensitivity analysis}
Uncertainty in the ISR2D model response is due to the model stochasticity and uncertainty in three model parameters. The ranges of the uncertain parameters are shown in Table~\ref{tab:uncertain_para}. These three uncertain inputs were assumed to be uniformly distributed within the given ranges. The model output of interest was the neointimal area as a function of time after stenting and its uncertainty was estimated using the non-intrusive method (NI) and the semi-intrusive (SI) method with surrogates of the blood flow micro model. We compared the uncertainty quantification result and sensitivity analysis result for similar values of UQ speedup. We also show the quasi-Monte Carlo (qMC) result from \cite{Nikishova2018} as the reference solution. The total number of model runs in both qMC and SI experiments was 1024.


\begin{table}[]
\centering
\begin{adjustbox}{width=0.8\textwidth}
\begin{tabular}{c|c|c|c}
\toprule
Uncertain Parameter           & Range (min) & Range (max) & Unit \\ \hline
inlet blood flow velocity     & 0.432       & 0.528       & m/s  \\ 
maximum deployment depth      & 0.09        & 0.13        & mm   \\ 
endothelium regeneration time & 15          & 23          & days \\ 
\bottomrule
\end{tabular}
\end{adjustbox}
\caption{List of UQ parameters of ISR2D and their min and max values.}
\label{tab:uncertain_para}
\end{table}

\begin{figure}[b!]
\centering
 \includegraphics[width=1.\textwidth]{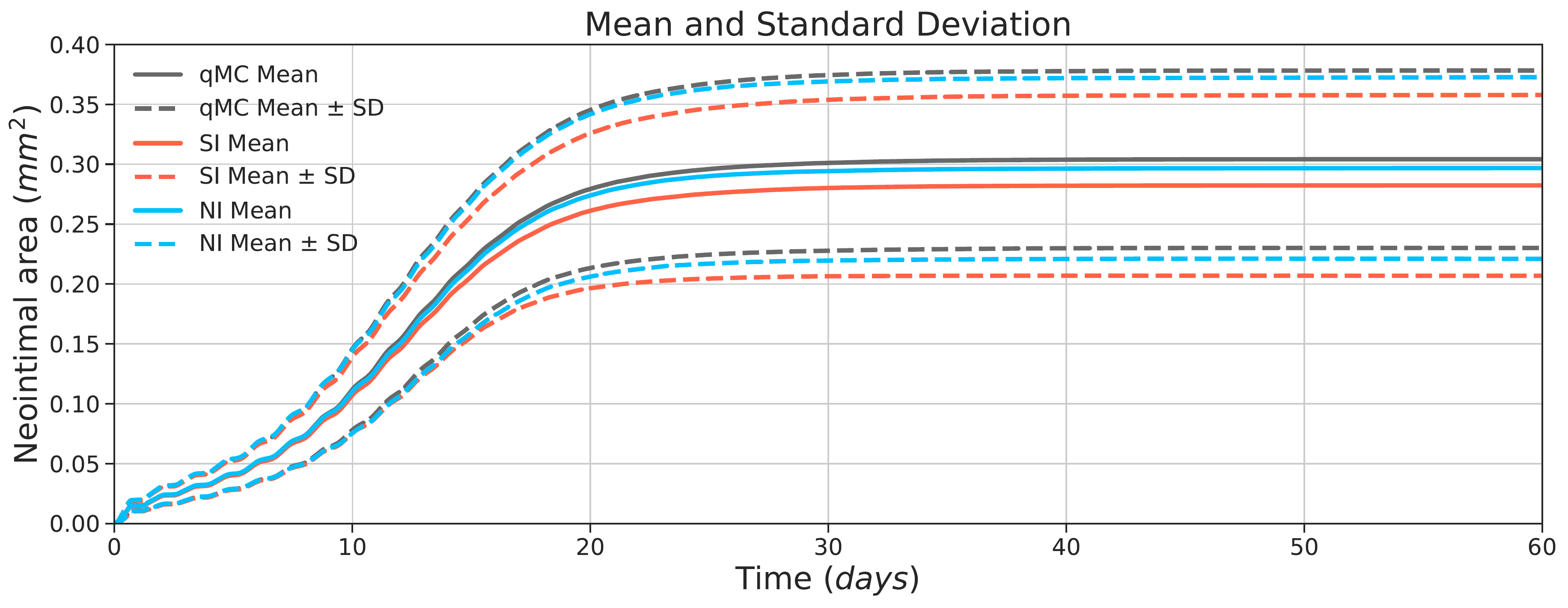}
 \caption{Mean and standard deviation of the ISR2D model output on the neointimal area with quasi-Monte Carlo (qMC) and with the semi-intrusive (SI) method and non-intrusive (NI) method.}
\label{fig:UQcomparisom_mean}
\end{figure}

Figure~\ref{fig:UQcomparisom_mean} shows the estimates of the mean and standard deviation with the qMC, the SI and NI methods. The mean is approximated especially well for the first 10 simulated days. After this point, slightly less average growth is observed in both SI and NI estimates than in the qMC results. The NI estimates slightly outperformed the SI estimates. The shape of the mean value of the neointimal area is well approximated by both SI method and NI method. The results of the standard deviation from the SI and NI methods also show approximately similar value to the qMC estimator.

\begin{figure}[tb!]
\centering
 \includegraphics[height=0.85\textheight]{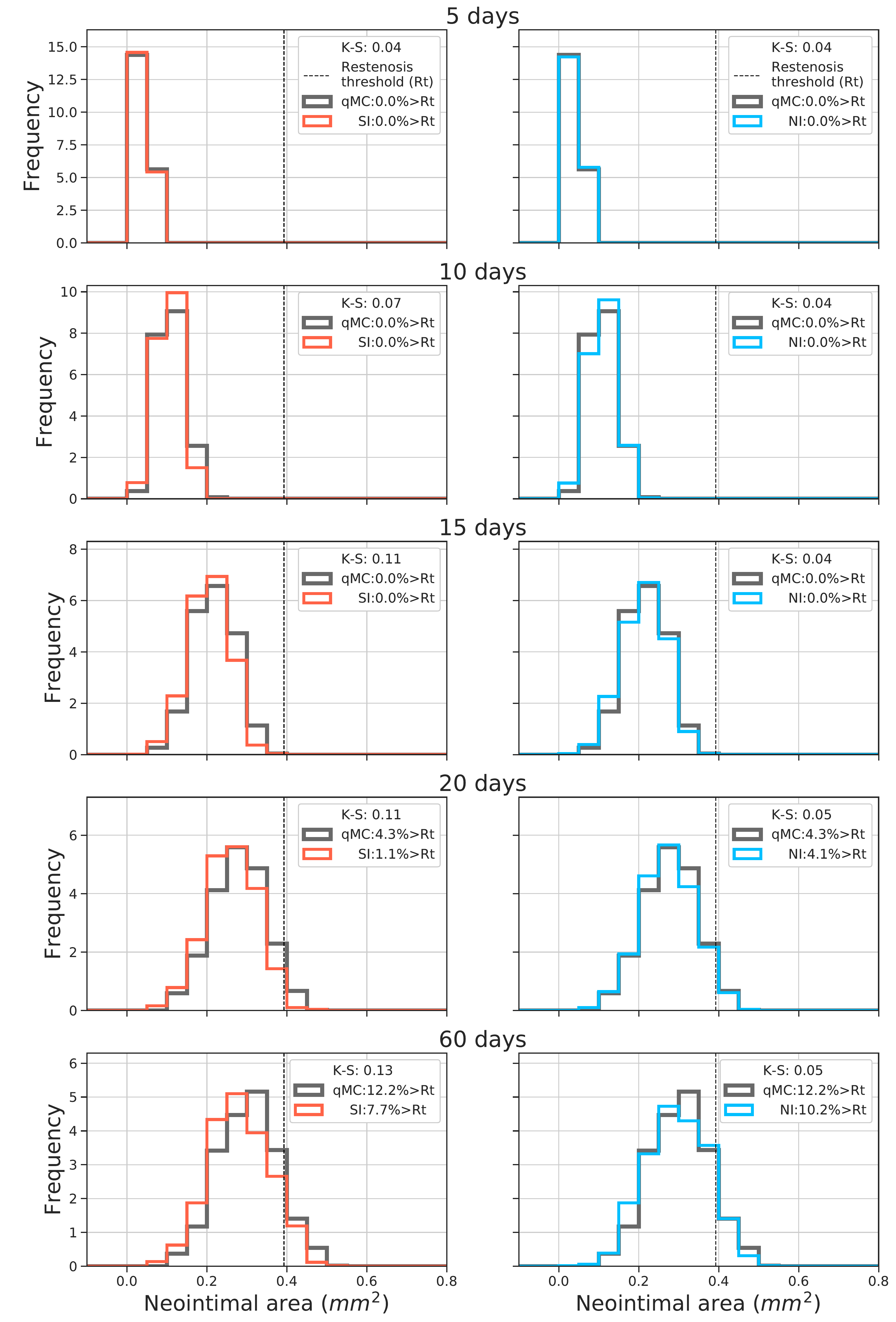}
 \caption{Histogram of the neointimal area at different simulation times obtained with the quasi-Monte Carlo (qMC), the semi-intrusive method (SI) and non-intrusive method (NI).}
\label{fig:UQcomparisom_pdf_time}
\end{figure}

The comparison of the histogram of obtained result with three UQ methods at 5, 10, 15, 20 and 60 days after stenting are shown in Figure~\ref{fig:UQcomparisom_pdf_time}. A good fit of the histograms is obtained at the early time points. For day 5 and day 10, the two sample Kolmogorov–Smirnov (K-S) test for qMC and SI distributions produces statistics of 0.04 and 0.07 respectively, while qMC and NI give  0.05 in both cases. At later time points, the K-S statistic is always smaller for the results of NI estimates than with SI estimates, which means a better fit. These plots also indicate the ratio of restenosis cases defined by 50\% occlusion of the original lumen area \cite{Serruys2009} (shown as the vertical line). Non-zero values of this ratio are only observed at day 20 and 60. 
As expected, prediction of a smaller growth in SI resulted in a relatively smaller predicted binary restenosis rate. The NI's prediction is closer to the qMC result. At the final simulation time point, NI predicted 10.3\% restenosis occurrence while SI predicted 7.7\% occurrence. 
A summary of uncertainty estimation at the final time step by different methods is presented in Table~\ref{tab:uq_comp}. The NI estimations have the smallest error in the estimation of the mean and the restenosis ratio. The SI with CNN results have a smaller error than some other methods for each estimator. All the SI and NI results show a statistically significant underestimation of the mean value (two-value t-test, $p < 0.01$). 

\begin{table}[tb]
\centering
\scriptsize
\begin{threeparttable}
\begin{tabular}{c|c|c|c|c|c|c|c}
\toprule
\multirow{2}{*}{UQ Method} &
\multirow{2}{*}{\bigcell{c}{Micro Model}} & 
\multicolumn{2}{c|}{\bigcell{c}{Mean Estimation\\ x 10$^{-1}$ ($\text{mm}^2$)}} & 
\multicolumn{2}{c|}{\bigcell{c}{ Standard Deviation \\x 10$^{-2}$ ($\text{mm}^2$)}} &
\multicolumn{2}{c}{\bigcell{c}{Restenosis Ratio\\(\%)}}  \\ \cline{3-8}
 & &Value&Error& Value&Error&Value&Error \\\hline
qMC &$\text{LBM}$\tnote{1}   &  3.04 & 0    & 7.41 &  0     & 12.2 & 0     \\ 
SI  &$\text{DD I}$\tnote{2}  &  2.88 & 0.16 & 7.43 &  0.02  & 8.3  & 3.9   \\
SI  &$\text{DD II}$\tnote{2} &  2.79 & 0.25 & 7.51 &  0.10  & 6.3  & 5.9  \\
SI  &$\text{Phys}$\tnote{2}  &  2.26 & 0.78 & 7.98 &  0.57 & 1.1  & 11.1 \\ 
SI  &CNN                     &  2.82 & 0.22 & 7.54 &  0.13 & 7.7  & 4.5   \\
NI  &/                       &  2.97 & 0.07 & 7.58 &  0.17 & 10.3 & 1.9  \\
\bottomrule
\end{tabular}
\begin{tablenotes}
\tiny{
\item[1] From \cite{Nikishova2018}.
\item[2] From \cite{Nikishova_2018_Semi-intrusive}.}
\end{tablenotes}
\end{threeparttable}
\caption{Comparison of the estimates of means and standard deviation of neointimal growth and restenosis ratio with qMC, SI and NI methods. The indicated error is the absolute difference from the reference qMC value. The four surrogate models for SIUQ are data-driven model I (DD I), data-driven model II (DD II), physics surrogate model (Phys) and convolutional neural network model (CNN). See \cite{Nikishova_2018_Semi-intrusive} for details on the Phys, and DD I and DD II surrogates.}
\label{tab:uq_comp}
\end{table}

\begin{figure}[tb!]
\centering
 \includegraphics[width=.90\textwidth]{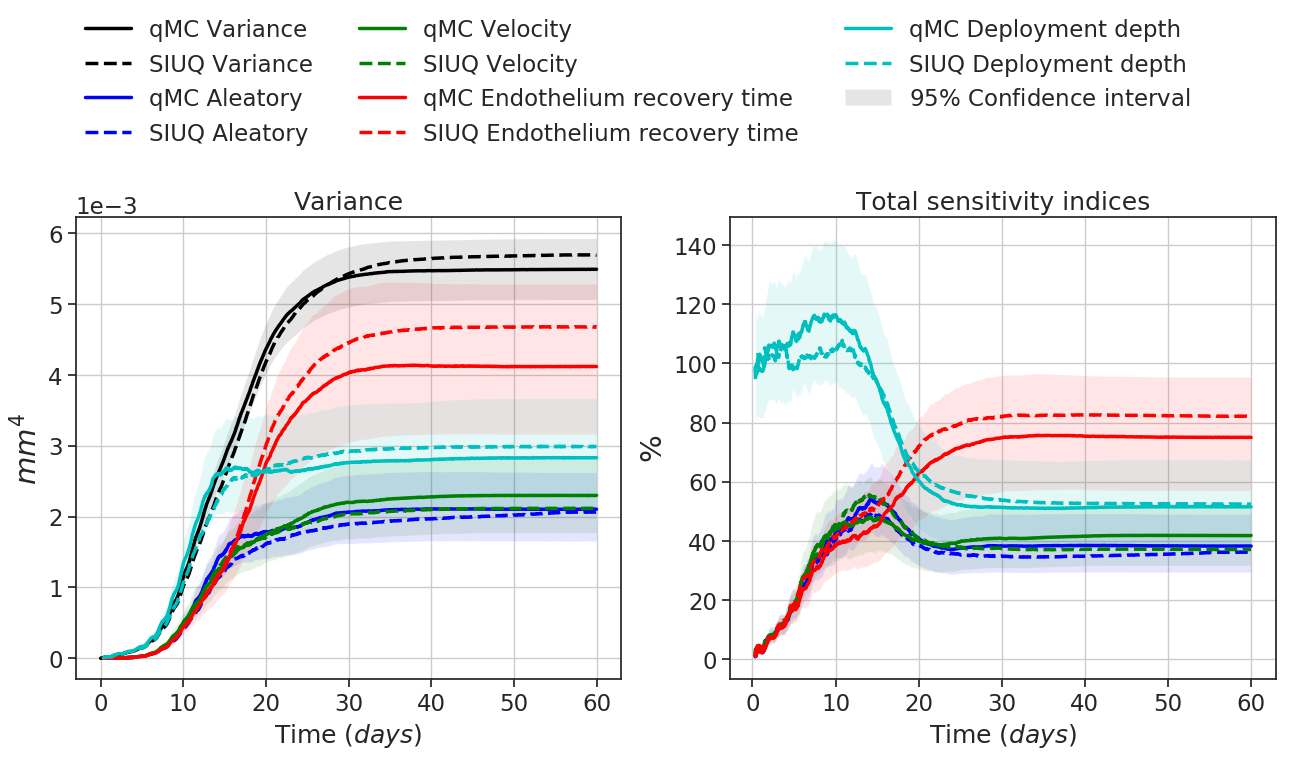}
 \includegraphics[width=.90\textwidth]{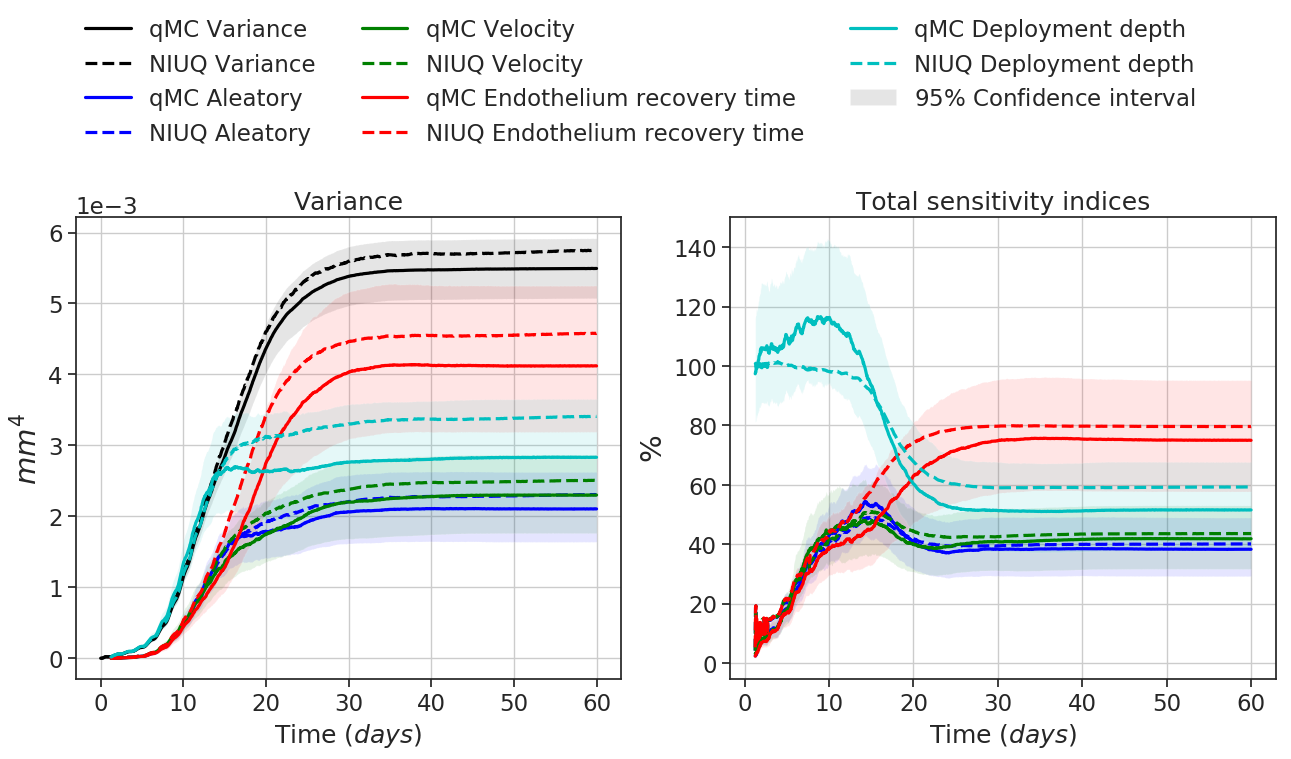}
 \caption{Partial and total variances (left column) and total Sobol sensitivity indices (right column) with qMC in solid lines and the SI (top row) and NI (bottom row) in dashed lines. Each of the quantities for the uncertain inputs includes the aleatory uncertainty. The area around the qMC results is the 95\% confidence interval obtained by bootstrap \cite{Archer_1997}.}
\label{fig:SAcomparisom}
\end{figure}

Figure~\ref{fig:SAcomparisom} illustrates the overall and the partial variances and Sobol sensitivity indices as quantified using SI, NI and qMC. These quantities were overestimated to a certain degree by the NI method but all the estimates are still within the confidence interval of the qMC results and the order of the partial variances is preserved. Since both the overall and the partial variances are overestimated, the error is significantly smaller in the estimation of the sensitivity indices, and the indices obtained by the NI method with GP are close enough to the one estimated by qMC. The variances estimated by the SI method are also close to the qMC result and all within its confidence interval. As a result, the total sensitivity indices are also well approximated with this type of method.

\subsection{Speedup}

\begin{table}[t!]
\centering
\scriptsize
\begin{threeparttable}
\begin{tabular}{c|c|c|c|c|c|c|c|c}
\toprule
\multicolumn{1}{c|}{\bigcell{c}{UQ\\ Method}} &
\multicolumn{1}{c|}{\bigcell{c}{Micro\\Model}} & \multicolumn{1}{c|}{\bigcell{c}{$\mathcal{T}_{\text{ISR}}$\\(min)}} & 
\multicolumn{1}{c|}{\bigcell{c}{$\mathcal{T}_{\text{ISR}^*}$\\(min) }} &
\multicolumn{1}{c|}{\bigcell{c}{$\mathcal{T}_{\text{micro}}$\\(min)}} &
\multicolumn{1}{c|}{\bigcell{c}{$\mathcal{T}_{\text{train}}$\\(min)}} & \multicolumn{1}{c|}{\bigcell{c}{$\mathcal{T}_{\text{sample}}$\\(min)}} & \multicolumn{1}{c|}{\bigcell{c}{N}} &
\multicolumn{1}{c}{\bigcell{c}{Speedup\\of UQ}} \\\hline
qMC &$\text{LBM}$\tnote{1}  & 89.4 & / & 74.9    & /     & /     &1024 &  1\\ 
SI  &$\text{DD I}$\tnote{2} & / & 50.9 & 37.8  & /     & 894     &1024 & 1.72\\
SI  &$\text{DD II}$\tnote{2} & / & 14.6 & 2.05  & /     & 447     &1024 & 5.94\\
SI  &$\text{Phys}$\tnote{2} & / & 11.9 & 0.08 & /   & /     &1024 & 7.51\\ 
SI  &CNN          &   /            &   12.8            &  0.26     & 9.9 & 447  &1024   &6.75\\
NI  &/                 & / & 0.17       & /      & 4.3 & $1.1\times 10^4$   &1024  & 6.82\\
\bottomrule
\end{tabular}
\begin{tablenotes}
\tiny{
\item[1] From \cite{Nikishova2018}.
\item[2] From \cite{Nikishova_2018_Semi-intrusive}.}
\end{tablenotes}
\end{threeparttable}
\caption{Comparison of the computational time and corresponding speedup of different approaches. The time value indicates the mean computational time obtained over $N=1024$ samples. $\mathcal{T}_{\text{micro}}$ is the execution time of the micro model (LBM/surrogate models) in one ISR2D simulation.
The computations were performed on the Distributed ASCI Supercomputer DAS5 \cite{7469992} with Intel Haswell E5-2630-v3 CPU.}
\label{tab:time_comparison}
\end{table}

In Table~\ref{tab:time_comparison}, the execution times and resulting speedups of the SI and NI methods relative to the qMC method are evaluated, including previously reported SI results from \cite{Nikishova_2018_Semi-intrusive} for comparison. Because of the light surrogate model, the SI approach with CNN was approximately seven times faster than black-box qMC, an improvement of more than a factor three over the nearest-neighbour interpolation based surrogate model. The simplified physics model was even faster, but was also the least accurate one, while the SI with CNN based surrogate model provided the best uncertainty quantification and sensitivity estimates among the four surrogates (see \cite{Nikishova_2018_Semi-intrusive} for details on the Phys, and DD I and DD II surrogates).

\subsection{Convergence}

\begin{table}[th]
\centering
\scriptsize
\begin{threeparttable}
\begin{tabular}{c|c|c|c|c|c|c|c}
\toprule
\multirow{2}{*}{\bigcell{c}{UQ \\ Method}} &
\multirow{2}{*}{\bigcell{c}{Micro \\ Model}} &
\multirow{2}{*}{\bigcell{c}{NMAE}} &
\multicolumn{2}{c|}{\bigcell{c}{Mean Estimation\\ x 10$^{-1}$ ($\text{mm}^2$)}} & 
\multicolumn{2}{c|}{\bigcell{c}{ Standard Deviation \\x 10$^{-2}$ ($\text{mm}^2$)}} &
\multirow{2}{*}{\bigcell{c}{KL Divergence \\ x 10$^{-2}$}}  \\ \cline{4-7}
 & & &Value&Error& Value&Error \\\hline
qMC &$\text{LBM}$\tnote{1}  & 0&  3.04 & 0    & 7.41 &  0     & 0     \\ 
\hspace{0.06cm} SI-$1^*$  &CNN & 6.06\%&  3.12 & 0.08 & 11.43 &  4.02  & 5.46    \\
SI-1  &CNN & 4.49\%&  2.85 & 0.19 & 7.87 &  0.46  & 2.98    \\
SI-2  &CNN & 3.89\%&  2.89 & 0.15 & 7.88 &  0.47  & 2.82    \\
SI-4  &CNN & 3.63\%&  2.82 & 0.22 & 7.54 &  0.13  & 2.59   \\
NI-20  &/ & /&  2.89 & 0.15 & 8.99 &  1.58 & 3.23     \\
NI-50  &/ & /&  2.92 & 0.12 & 7.28 & 0.13 & 2.82   \\
NI-150  &/ & /&  2.97 & 0.07 & 7.58 &  0.17 & 2.53   \\
\bottomrule
\end{tabular}
\begin{tablenotes}
\tiny{
\item[1] From \cite{Nikishova2018}.}
\end{tablenotes}
\end{threeparttable}
\caption{Comparison of the estimates of means and standard deviation of neointimal growth and Kullback–Leibler divergence with qMC, SI and NI methods using different training sample size. SI/NI-X denotes a semi-intrusive or non-intrusive UQ with a surrogate model trained by the data from X runs of ISR2D simulation. SI-$1^*$ denotes a semi-intrusive UQ with a surrogate model trained by the truncated data of one simulation. The third column shows the NMAE of the surrogate model of the blood flow simulation on the validation dataset. The last column shows the relative entropy of each output distribution compared to the output distribution of qMC result.}
\label{tab:uq_conv}
\end{table}

A study of the effect of training sample size on convergence of SIUQ and NIUQ is shown in Table~\ref{tab:uq_conv}. The mean estimation, standard deviation and Kullback–Leibler (KL) divergence \cite{KLD1951,Kullback59} are estimated for the distributions at the last time step. The four semi-intrusive UQ results are based on the surrogate models trained with the data from 1,2, and 4 ISR2D simulations respectively. SI-$1^*$ denotes a case in which the surrogate model is trained with only part of the data of one ISR2D simulation (the truncated data from day 0 to day 15). The error in standard deviation is reduced significantly with improvement of the surrogate model, however the tendency in mean estimation is not obvious. In SI-$1^*$, the surrogate model was trained with truncated data of one simulation, which resulted in an overestimation of mean. When the surrogate model was trained on complete data of one simulation or more, the mean estimates are all shifted to minor underestimations around 0.285 $\text{mm}^2$. The KL divergence shows that the output distribution of the SI models with a better CNN surrogate model has a lower relative entropy compared to the qMC result, which means a better output distribution approximation. It is important to note that in semi-intrusive UQ, the quality of a surrogate model influences the uncertainty estimation in a complicated way. The output of the surrogate model is intermediate information in between submodels, and the corresponding error introduced by the surrogate model might be retained, alleviated or aggravated through other computations and further iterations in a multiscale simulation. Therefore, the influence of the quality of the surrogate model on the convergence of the semi-intrusive method will be different for each multiscale model.  

A significant improvement of uncertainty estimates was found when the training sample size of the GPR surrogate model was increased from 20 to 50, especially for the standard deviation. However, adding even more training data had limited impact on the uncertainty estimates, as can be observed by comparing the results for NI-50 and NI-150. The surrogate model of non-intrusive UQ is relative easier to control as the QoIs are directly mapped from uncertain inputs. Typically, the more information one has, the better surrogate model can be achieved.

\section{Discussion}\label{sec:Discussion}

The CNN surrogate model performed well regarding the wall shear stress prediction for the micro model. It takes advantage of convolution layers to fetch latent features in the geometry input and then uses the FC layer and deconvolution layers to map the features to the wall shear stress prediction. 
Although the CNN surrogate model was able to predict the wall shear stress quite accurately, a small error still exists. This error introduced by the surrogate model then propagated through the iteration and led to the error in the uncertainty estimation and restenosis prediction as shown in Figure~\ref{fig:UQcomparisom_mean} and Table~\ref{tab:uq_comp}. The accuracy of the estimation with SIUQ method depends not only on the quality of the surrogate model but also on the structure of the multiscale simulation. However, the UQ result from the SI method suggests that the error is small enough to produce uncertainty and sensitivity estimates close to the ones obtained by qMC. Of course, the UQ result can be further improved by a better CNN surrogate model, e.g by training with a larger dataset or constructing a deeper CNN structure. But such improvement in the surrogate model does not necessarily guarantee a significant improvement in the uncertainty estimations. 
A trial experiment has been performed on ISR2D with a better CNN surrogate model (NMAE $\approx 1\%$), but the improvement of the corresponding uncertainty estimates was minuscule. 

The UQ estimates with CNN surrogate model outperformed the result of most previous surrogate models except DD I. The result with DD I is still slightly better which may be due to the large training dataset it used \cite{nikishova2019semi}. However the obtained speedup of DD I is much lower than the CNN model, since the CNN model learned the latent pattern of the data, while DD I simply looked for similar cases among all the training data. Because of this, the prediction cost of CNN is significantly lower. 
The maximum expected speedup of SIUQ with a surrogate model is limited by the computational cost of the macro model in the multiscale simulation. It can be calculated by $\mathcal{T}_{\text{ISR}}/\mathcal{T}_{\text{macro}}$, which is around six times for ISR2D. However the speedup of the SIUQ with CNN surrogate model (Table~\ref{tab:time_comparison}) is even higher than the maximum expected value. This is because the computational cost of the SMC (macro) model varies. It mainly depends on the number of agents in the model. As the ISR2D with a surrogate model underestimates the neointimal growth which means fewer agents in the SMC model, the corresponding computational time is reduced from 14 minutes to 12 minutes and even less. In the speedup calculation, $\mathcal{T}_{\text{sample}}$ includes the data generation time for both training data and validation data. The computational cost of this part can be further reduced by applying a cross validation method, such as k-fold cross validation. The cross validation requires no additional dataset but validates predictions within the training data. However, the improvement is limited, as the computational burden of the SIUQ method mainly comes from the cost of the simulation multiplied by the number of samples. When the number of samples required in UQ or sensitivity analysis is large ($N$ in Equation~\ref{eq:speedup} tends to infinity), the speedup will not be affected visibly by the cost of obtaining the surrogate model. Although the deployment of the surrogate model reduces the computational cost, the maximum speedup is still limited by the computational cost of the other parts of the multiscale simulation. 

In NIUQ, these other parts are also emulated by the surrogate model, with the speedup converging to the ratio between the time taken to run the original model and the time taken to evaluate the surrogate as $N$ goes to infinity. For finite $N$, the total time taken by the surrogate-based model may be dominated by the collection of training data, and minimizing the required amount of data (e.g. by applying GPR with active learning \cite{activelearning2015,jones1998efficient}) may substantially improve performance. Of course, the exact amount of data required to train a surrogate model for NIUQ depends not only on the dimensionality of the input but also on the desired accuracy. Therefore, selecting which method to apply for achieving an efficient UQ analysis has to take into consideration the multiscale model itself and the requirements of the UQ analysis.


By comparing the SI and NI results at the similar computational efficiency, one can see that uncertainty and sensitivity analysis of SI were as good as the NI. SI has the additional advantage of granularity, since only part of the model is replaced by the surrogate. This means that the parameters of the submodels not replaced by the surrogate can be varied and studied without changing the surrogate, as long as the replaced micro model is not affected. For example, in case of the ISR2D model, different parameters and rule sets for cell behaviour can be used with the existing surrogate model for flow. On the other hand, using a NI model for a different biological ruleset would require essentially building a new NI surrogate, which would incur a significant computational cost.

In general, both SI and NI approaches performed well. The SI approach is more suitable for cyclic multiscale simulations as it retains the framework of the simulation and can obtain the training data for the surrogate model at a relatively low cost. Another semi-intrusive UQ strategy is to run the simulations for UQ and build the surrogate model on the fly. Such a dynamic system should be capable of constructing and updating the surrogate during the simulation process, for instance, as done by Leiter et al. \cite{LEITER201891}. In the UQ scenario, the surrogate model can also be validated dynamically \cite{nikishova2019semi}. We aim to apply these techniques to the three-dimensional versions of the ISR model in future work.

\section{Conclusions and future work}\label{sec:Conclusion}
In order to implement uncertainty quantification and sensitivity analysis efficiently for the ISR2D model, a new surrogate model based on a convolutional neural network was developed and applied in semi-intrusive uncertainty quantification method.
The UQ estimate with the new surrogate model was compared with the result of previous work and a non-intrusive UQ estimates based on Gaussian process regression. The result shows that SI with the convolutional neural network surrogate model outperformed the previously developed semi-intrusive surrogate models. The result is also comparable to non-intrusive estimates. Both SI and NI are valid methods to perform UQ in an efficient way for the ISR2D model. 

In this study, we applied standard quasi Monte Carlo method with a surrogate model for both non-intrusive and semi-intrusive uncertainty quantification studies. The surrogate model in both cases will inevitably introduce errors into uncertainty estimation. A multi-fidelity Monte Carlo method \cite{MFMC2016,MFMC2018} can be applied here for uncertainty analysis as both methods require to run a certain amount of original simulation to generate training data. Therefore a multi-fidelity framework can combine both high fidelity data (training data) and low fidelity approximation (generated by or with a surrogate model) together to further improve the accuracy of uncertainty estimation. This method has been widely used in non-intrusive UQ analysis, but hasn't been applied to semi-intrusive UQ methods yet. It would be helpful to include the method to our further works.

In this study, we have taken ISR2D simulation as a case study and investigated how three uncertain parameters: blood flow velocity, endothelium regeneration time and deployment depth affect the neointimal area of restenosis. However, ISR2D is a simplified model for the restenosis process. We aim to apply the UQ techniques to our more realistic and complex model, ISR3D \cite{Zun2017,Zun2019}, and analyse uncertainty parameters including not only the three mentioned in this paper but also other factors, such as fenestration percentage, maximum strain for SMCs and others that may influence the growth. Since ISR3D is suitable for modelling restenosis in realistic geometries, it can eventually be used to perform in silico clinical trials or to design novel stents. These directions of research would require many runs of the 3D model, and this makes it essential to run each individual simulation as cheaply as possible, while retaining the validity of the model. The 3D flow calculation is much more expensive in 3D; for example, the 3D simulations of restenosis done for the InSilc project\footnote{https://insilc.eu/} take around 3000 core hours for each artery reconstructed from optical coherence tomography (OCT) images. Substituting the flow model with a surrogate may be essential for reducing these costs.

Semi-intrusive methods for UQ of multiscale models require modifying those models by adding new submodels or other components, and by changing the connections between them. If the codes are tightly coupled using a low-level communication facility such as MPI, then this entails changing the model code and maintaining multiple versions simultaneously, which is cumbersome and error-prone. We have recently created a new implementation of the multiscale coupling framework, MUSCLE3 \cite{Veen2020,MUSCLE3_032}. With MUSCLE3, submodels are connected to the framework once and can then be coupled in different ways based on a simple configuration file. Advanced non- and semi-intrusive UQ algorithms can be implemented as additional modules, and wired into the multiscale model without changing the code of the submodels. We have already ported ISR3D to MUSCLE3, and aim to implement non- and semi-intrusive UQ with MUSCLE3.

\section{Funding}
This work was supported by the Netherlands eScience Center under grant agreement 27015G01 (e-MUSC project). This project has received funding from the European Union Horizon 2020 research and innovation programme under grant agreements \#800925 (VECMA project) and \#777119 (InSilc project). PZ has received funding from The Russian Foundation for Basic Research under agreement \#18-015-00504 and from the Russian Science Foundation under agreement \#20-71-10108. This work was sponsored by NWO Exacte Wetenschappen (Physical Sciences) for the use of supercomputer facilities, with financial support from the Nederlandse Organisatie voor Wetenschappelijk Onderzoek (Netherlands Organization for Science Research, NWO).

\bibliography{main}

\newpage

\end{document}